\documentclass[aps,pra,twocolumn,superscriptaddress,longbibliography]{revtex4-1}

\usepackage{natbib}
\usepackage{pgfplots}
\usepackage[breaklinks]{hyperref}
\usepackage{amssymb}
\usepackage{amsmath,amsthm}
\usepackage{amsfonts,dsfont}
\usepackage{graphicx}
\usepackage{lipsum}
\usepackage{subcaption}
\usepackage{mathtools}
\usepackage{tikz}
\usepackage{cleveref}
\usetikzlibrary{positioning}

\usepackage[normalem]{ulem}
\usepackage{color}
\usepackage{xcolor}

\newcommand{\cev}[1]{\reflectbox{\ensuremath{\vec{\reflectbox{\ensuremath{#1}}}}}}

\newcommand{\rr}{{\mathbf r}}
\newcommand{\zz}{{\mathbf z}}

\newcommand{\pp}{\mathbf{p}}

\newcommand{\F}{\mathcal{F}}        
\newcommand{\E}{\mathcal{E}}        

\newcommand{\Tr}{\mbox{Tr}}

\newcommand{\bra}[1]{\mbox{$\langle #1 |$}}
\newcommand{\ket}[1]{\mbox{$| #1 \rangle$}}

\newsavebox{\mstrut}
\newcommand{\bbra}[1]{%
    \sbox{\mstrut}{\(#1\)}%
    \mathinner{\left\langle\kern-0.5\ht\mstrut\left\langle{#1}\right|\mkern-2mu\right|}%
}
\newcommand{\kett}[1]{%
    \sbox{\mstrut}{\(#1\)}%
    \mathinner{\left|{#1}\right\rangle\kern-0.5\ht\mstrut\rangle}%
}

\begin{document}
\title{Orbital-Free Quasi-Density Functional Theory}

\author{Carlos L. Benavides-Riveros} 
\email{cl.benavidesriveros@unitn.it}
\affiliation{Pitaevskii BEC Center, CNR-INO and Dipartimento di Fisica, Università di Trento, I-38123 Trento, Italy}

\date{\today}

\begin{abstract}
Wig\-ner functions are broadly used to probe non-classical effects in the macroscopic world. Here we develop an \textit{orbital-free} functional framework to compute the 1-body Wig\-ner quasi-probability for both fermionic and bosonic systems. Since the key variable is a quasi-density, this theory is particularly well suited to circumvent the problem of finding the Pauli potential or approximating the kinetic energy in orbital-free density functional theory. As proof of principle, we find that the uni\-ver\-sal functional for the building block of optical lattices results from a translation, a con\-trac\-tion, and a rotation of the corresponding functional of the 1-body reduced density matrix, indicating a strong connection between these functional theories. Furthermore, we relate the concepts of \textit{Wig\-ner negativity} and \textit{$v$-representability}, and find a manifold of ground states with negative Wig\-ner functions.  
\end{abstract}

\maketitle

\textit{Introduction.---} Detecting and understanding quantum features at the macroscopic level is one of the main theoretical and technological challenges of modern quantum sciences. Nowadays, state-of-the-art experiments can directly observe non-classical behavior (as quantum superposition) in systems with a truly macroscopic number of particles, with as many as $10^{16}$ atoms \cite{Friedman2000,Fein2019,Wollack2022,PhysRevLett.130.133604}. A powerful theoretical and computational strategy to detect that \textit{quan\-tumness} is by directly measuring the sys\-tem's corresponding Wig\-ner function. Although normalized to uni\-ty, Wig\-ner functions are quasi-probability distributions that can take negative values, a phenomenon that has no classical counterpart. Hence, negativity in the Wig\-ner functions has been linked to non-classical features of quantum states and is considered a distinctive signature of quantum entanglement \cite{PhysRevLett.119.183601,entan,AnatoleKenfack2004}, contextuality \cite{PhysRevLett.101.020401,PhysRevA.95.052334,Howard2014}, quantum computation \cite{Veitch_2012}, quantum steering \cite{PRXQuantum.1.020305,walschaers2022quantum}, or even quantum gravity \cite{PRXQuantum.2.010325}.

Due to the exponentially large Hilbert spaces of quan\-tum ma\-ny-body systems, finding the corresponding Wig\-ner function is, in general, a computationally prohibitive task. Yet for identical particles it is possible to circumvent the Hilbert space's exponential growth by means of a universal functional of certain re\-du\-ced, more manageable, quantities, like, e.g., the  density. Based on the important observation that electronic systems are fully determined by the ground-state density \cite{HK}, density functional theory (DFT) is a prominent methodology in electronic structure calculations, with applications ranging from quantum chemistry and ma\-te\-rial science \cite{Jones15, D2CP02827A} to self-driving labs \cite{huang2023selfdriving}. Quite remarkably,  or\-bi\-tal-free DFT achieves a computational linear  scaling with the system size \cite{Ryczko}. But, unfortunately, from the density alone it is not possible to reconstruct the Wig\-ner function, and therefore standard DFT is, in general, not suitable for describing non-classical features of quantum many-body systems.

A recent phase-space formulation of DFT employs, as the central variable, the 1-particle Wig\-ner quasi-den\-si\-ty, which is in a one-to-one correspondence  with the res\-pec\-tive ground state for interacting ma\-ny-fermion/boson systems \cite{BGBV}. Its main feature is that the 1-bo\-dy Wig\-ner function can be accessed directly, without pre-computing the full wave function. This Wig\-ner quasi-density functional theory (qua\-si-DFT) is a promising theoretical tool to model many-body problems while accounting for non-classical features, strong interactions, and quantum correlations, with the same computational cost as standard DFT. As we will show below, the theory has also the potential of bypassing well-known problems of orbital-free DFT. To date, however, there are neither orbital-free nor orbital-dependent functionals for quasi-DFT. 

Here, we will obtain equations for the fer\-mio\-nic/bo\-so\-nic 1-particle qua\-si-density. This is, we will argue, the initial step to developing a full \textit{orbital-free} framework for Wigner qua\-si-DFT. As one of our main results, we will show that $\omega(\rr,\pp)$, the 1-particle Wig\-ner quasi-density, satisfies the following, effective, eigen-equation:
\begin{align}
h_{\rm eff} \star \omega(\rr,\pp) = \omega(\rr,\pp) \star h_{\rm eff}  = \mu \, \omega(\rr,\pp)\,,
\label{eq.crucial}
\end{align}
where $h_{\rm eff}= \tfrac12\pp^2 + v_{\rm ext}(\rr) + v_{\rm eff}(\rr,\pp)$, $v_{\rm ext}(\rr)$ is the external potential,  $v_{\rm eff}(\rr,\pp)$ is certain effective potential that we introduce below, and the symbol $\star$ is the so-called star product of phase-space quantum mechanics. 

The letter is organized as follows: First, we review both the orbital-free functional theories and the Wig\-ner formulation of DFT. Second, we derive an Euler-Lagrange equation for the 1-body Wig\-ner quasi-density. Next, we derive an equation using the Moyal product. We then employ the Hubbard model to present for the first time a functional realization of quasi-DFT. We conclude with a summary and discuss some implications of our results. In the Appendixes, we provide additional technical details. 

\textit{Functional theories.---} The enormous success of DFT in electronic structure calculations is mainly due to the existence of a set of self-consistent 1-particle equations that allow for the computation of the density from 1-particle orbitals \cite{KS}. Although it is much cheaper than wave-function methods, this Kohn-Sham DFT still has an unfavorably computational scaling with the cube of the number of electrons \cite{engel2011density}. In turn, orbital-free DFT allows a much more favorable, linear scaling with the system size \cite{D2CP02827A,Ryczko}, but this computational advantage is counterbalanced by the fact that the quantum mechanical kinetic energy functional is unknown, and it is written as a classical, approximate function of the electron density. A parallel intellectual effort is the 1-particle reduced density matrix functional theory (1-RDMFT) that exploits the full 1-particle picture of the many-body problem by seeking a universal functional of the 1-body reduced density matrix  (1-RDM),  \cite{Gilbert, DP, Pernal2016}, for fermionic \cite{Schilling2018,PirisPRL,PhysRevB.99.224502}, bosonic \cite{PhysRevLett.124.180603,PhysRevResearch.3.013282,Maciazek_2021,Liebert_2023}, or relativistic  \cite{10.21468/SciPostChem.1.2.004} interacting particles. Similar to DFT, 1-RDMFT is based on a one-to-one correspondence between the ground state and its corresponding 1-RDM. Although this theory is in a better position than DFT to tackle strong correlations \cite{Sharma}, its broad use has been ham\-pe\-red by the absence of Kohn-Sham-like equations for the natural orbitals (i.e., the eigenvectors of the 1-RDM) \cite{Pernal2016, PernalPiris}. It, therefore, comes as no surprise that fermionic 1-RDMFT is computationally much more demanding than DFT \cite{PNAS}. Unfortunately, there are quite a few orbital-free formulations of 1-RDMFT (the most notable being the exchange part of the Hartree-Fock functional). The development of an orbital-free perspective of 1-RDMFT could boost its broad applicability.

\textit{Phase-space quantum mechanics.---} In the phase-space formulation of quantum mechanics, observables are represented by symbols, i.e., functions of position $\rr$ and momentum $\pp$ coordinates. Out of many choices, Wig\-ner functions host the most natural representation of quantum mechanics \cite{Wigner}. In the classical limit, it turns out to be the  phase-space distributions of statistical mechanics \footnote{There are other phase-space distributions  (e.g.,~Berezin's $Q$ or $P$ functions), but the  Wig\-ner functions are the only ones that are real and give the correct marginal probabilities.}. In this formulation, quan\-tum operators correspond uniquely to phase-space classical  functions via the Weyl correspondence, while ope\-ra\-tor products correspond to $\star$-products. This noncommu\-ta\-ti\-ve star (twisted or Moyal) product is commonly defined by the phase-space pseudo-differential operator \cite{moyal_1949}:
$\star \equiv \exp [i\hbar(\cev\partial_r\vec{\partial}_p -
\cev\partial_p\vec\partial_r)/2]$; the arrows denote that a given derivative acts only on a function standing on the left/right. This product is defined by $\mathcal{Q}(f\star g) =  \mathcal{Q}(f)\mathcal{Q}(g)$ where $\mathcal{Q}(f)$ is the  quantized operator version (by the Weyl rule) of the phase-space  function $f$ \cite{Weyl1927}. The eigenvalue problem for the Hamiltonian $H$  reads $H \star × f_n = E_n f_n = f_n \star H$ \footnote{Another important property  is that the integral of the $\star$-mul\-tiplication reduces to a plain integral, namely: $\int d\Omega \, (f \star g) = \int  d\Omega \, (g \star f) =  \int d\Omega fg$, with $d\Omega = d^3r d^3p$.}.

\textit{1-body Wig\-ner quasi-density.---} By definition, the 1-bo\-dy Wig\-ner quasi-densities are given in terms of the 1-RDM $\gamma(\rr,\sigma;\rr',\sigma')$, by the relation:
\begin{align}
\omega^{\sigma\sigma'}(\rr,\pp) 
= \frac{1}{\pi^{3}} \int \gamma(\rr - \zz, \sigma; \rr + \zz, \sigma')\,
e^{2i \pp\cdot \zz} \,d^3z\,,
\label{eq:Wigner} 
\end{align}
where $\sigma \in \{\uparrow,\downarrow\}$ are the spin variables. Notice that the marginal  $\sum_{\sigma}\int \omega^{\sigma\sigma}(\rr,\pp) \,d^3p$ gives exactly the density $n(\rr)$ (the central object of DFT) \footnote{To ease the no\-ta\-tion we skip from now on the spin indices in $\omega$.}.

\textit{Wig\-ner quasi-DFT.---} A generalization of the Ho\-hen\-berg-Kohn \cite{HK} and Gilbert \cite{Gilbert} theorems to Hamiltonians of the form $H = h + V$, with a \textit{fixed} two-particle interaction $V$, proves the existence of a universal Wig\-ner functional $\F_V[\omega]$ of the 1-body quasi-density $\omega$ \cite{BGBV}. Indeed, for any choice of the 1-particle phase-space Hamiltonian $h(\rr,\pp)=\frac12 \pp^2 + v_{\rm ext}(\rr,\pp)$ the energy functional:
\begin{align}
\E[\omega] \equiv   \int h(\rr,\pp) \omega(\rr,\pp) d\Omega +\F_V[\omega] \geq E_{\rm gs} \,,
\label{eq2}
\end{align}
is bounded from below by the exact ground-state energy.  The equality in Eq.~\eqref{eq2} holds exactly when $\E[\omega]$ is evaluated using the ground-state 1-body quasi-density $\omega_{\rm gs}$.  As in standard DFT, the functional $\F_V[\omega]$ is completely independent of any external (phase-space) potential $v(\rr,\pp)$. As in 1-RDMFT, it is also completely independent of the kinetic energy and depends only on the fixed two-particle interaction $V$. The (universal) functional $\F_V[\omega]$ obeys a constrained-search formulation, by considering only ma\-ny-body wave-functions that integrate to the same $\omega$:
\begin{align}
\F_V[\omega] = \min_{\Psi \rightarrow \omega} \bra{\Psi} V\ket{\Psi}\,.
\end{align}
While the functional is unknown, it is known that it has some better scaling properties than the functionals in DFT. For instance, by  defining, $\omega_\lambda =  \omega(\lambda\rr, \lambda^{-1}\pp)$, one can show that $\F_V[\omega_\lambda] = \lambda\F_V[\omega]$ \cite{BGBV}, a fact that lies in the exact knowledge of the kinetic energy functional. 

\textit{Representability condition of $\omega$.---} In a quite natural way, 1-body quasi-densities inherit the representability conditions of the 1-RDM, $\gamma$. Due to unitary invariance, those can be expressed as conditions on the eigenvalues of $\gamma$ \cite{RevModPhys.35.668}. Therefore, it is convenient to use the spectral representation of $\omega$ (i.e., $\omega = \sum_i n_i f_i$) to find its re\-pre\-sen\-tability conditions. In general, $n_i\geq 0$. In addition, in the case of fermions:
\begin{align}
\omega \star \omega \leq \omega\,,
    \label{rep}
\end{align}
which is just a consequence of the Pauli exclusion principle \footnote{There are in the literature other representability conditions for the spin coordinates \cite{PhysRevA.87.022118}}. 

\textit{Equation for the 1-body quasi-density.---} We now exhibit an exact equation for the phase-space 1-body quasi-density. Let $\E[\omega]$ be the energy functional of the Wig\-ner function \eqref{eq2}, subject to the constraint $\int d\Omega \,\omega(\rr,\pp) = N$. The $N$-particle phase-space density which minimizes such a functional is found by applying a functional derivative of the Lagrangian $\E[\omega] - \mu N$ with respect to $\omega$, yielding the Euler-Lagrange equation of Wigner quasi-DFT:
\begin{align}
\label{eq3}
h(\rr,\pp) + \frac{\delta\F_V[\omega]}{\delta \omega(\rr,\pp)} = \mu\,.
\end{align} 
There is an important consequence of this result. As is well known, one of the central problems in orbital-free DFT is approximating the kinetic energy functional in terms of the density \cite{PhysRevB.75.205122,PhysRevA.63.052508} or, alternatively, the Pauli potential \cite{doi:10.1063/1.4940035,PhysRevResearch.2.013159}. It is, indeed, particularly crucial that the Pauli principle be captured precisely in the kinetic energy. As we can see in Eq.~\eqref{eq3}, this important problem is completely absent in the  phase-space formalism. First, the kinetic energy and the external potential are exact, rather simple phase-space functions, and no approximation is needed. Second, the representability condition of the Wig\-ner function \eqref{rep} guarantees that the Pauli principle is fulfilled. As a consequence, our orbital-free quasi-DFT needs only to approximate the universal functional $\mathcal{F}_V[\omega]$.  

\textit{$\star$-eigenequation for $\omega$.---} Inspired by the work of Levy, Perdew and Sahni \cite{PhysRevA.30.2745} we exhibit now an exact $\star$-ei\-gen\-equation for the 1-particle quasi-density. As explained in Appendix \ref{app3}, by computing the directional functional derivative of $\mathcal{E}[\omega]$ at the point $\omega$ in the direction of $\omega$ one can show that $\omega_{\rm gs}$ fulfills the following equation:
\begin{align}
\label{eq.crucial}
\omega_{\rm gs} \star h_{\rm eff} = h_{\rm eff}\star \omega_{\rm gs} = \mu \, \omega_{\rm gs}\,,
\end{align}
where $h_{\rm eff}=h + \delta\F_V[\omega]/\delta \omega|_{\omega = \omega_{\rm gs}}$. The simplicity of this formula can be compared with the one from orbital-free DFT for  $\sqrt{n(\rr)}$ \cite{MARCH1986476}. Noticeably, the formula \eqref{eq.crucial} allows for a Wig\-ner-Moyal expansion of the equation for the qua\-si-density:
$\sum_n \frac{i^n\hbar^n}{2^n n!}
h_{\rm eff} (\cev\partial_r\vec{\partial}_p - \cev\partial_p\vec\partial_r)^n \omega = \mu \omega$.

\textit{Functional realization.---} To the best of our knowledge, there are no explicit functionals of $\omega$. Although  1-RDMFT functionals could be Wig\-ner transformed, almost all of them are written in terms of natural orbitals \cite{Piris2,doi:10.1063/5.0139897,PhysRevLett.127.233001,MULLER1984446,Giesbertz,Pernal,PhysRevLett.128.013001,PhysRevResearch.3.L032063,Gibney2022,PhysRevA.98.022504}, so they are not suited for our purposes. Let us, therefore, illustrate the potential of  orbital-free quasi-DFT by discussing the generalized Bose-Hubbard dimer, whose standard version has been broadly used to unveil aspects of functional theories \cite{PhysRevLett.124.180603,Cohen1,Schilling2019,burke2022lies,10.3389/fchem.2021.751054,10.1063/5.0143657}. The interacting Hamiltonian, containing all particle-conserving quartic terms, can be written with 3 parameters in the following way:
\begin{align}
V(u_1,u_2,u_3) &=  u_1 \sum_{j=l,r}\! \hat{n}_j(\hat{n}_j-1) \nonumber \\ &  + u_2 \hat{n}_l\hat{n}_r + u_3 \left[(b^\dagger_l)^2 b_r^2 + (b^\dagger_r)^2 b_l^2\right]\,,
\label{hamiltletter}
\end{align}
where $b^\dagger_{j}$, $b_{j}$ and $\hat n_{j}$ are the creation, anhilitation and par\-ti\-cle-number ope\-ra\-tors on the left/right sites $j\in\{l,r\}$. Normalizing to 1 and assuming  real-valued matrix elements, the 1-RDM can be represented in the lattice-site basis $\ket{l}$, $\ket{r}$ as
\begin{align}
\gamma = \left(\tfrac12+\vec{\gamma}\cdot \vec{\sigma}\right)\,,
\end{align}
where $\vec{\gamma} = (\gamma_{lr},0,\gamma_{ll}-\tfrac12)$, $\vec{\sigma}= (\sigma_{x},\sigma_y,\sigma_{z})$ is the vector of Pauli matrices, and $\gamma_{ij} = \bra{i}\gamma \ket{j}$. 

\begin{figure}[htb]
\begin{tikzpicture}
 \node (img) {
  \includegraphics[scale=0.3]{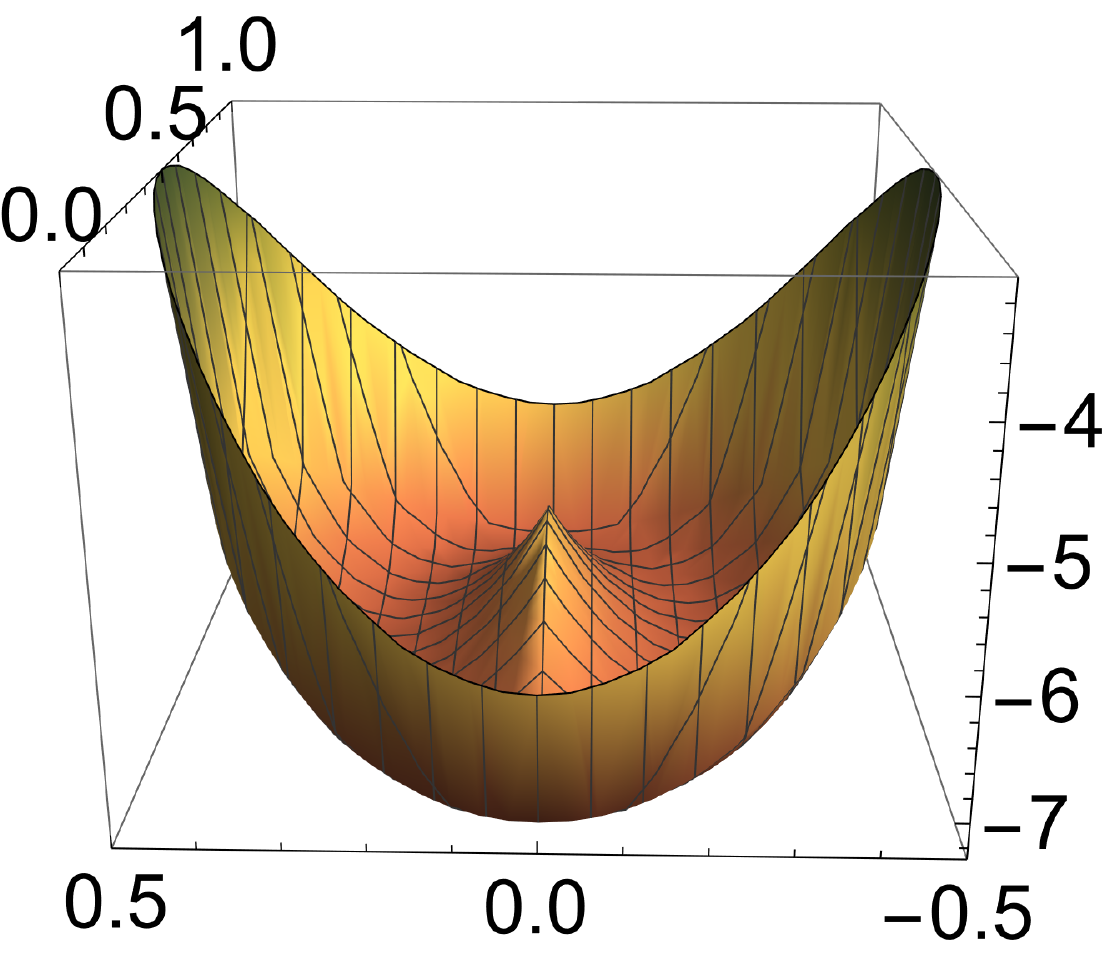}\hspace{0.7cm}
  \includegraphics[scale=0.3]{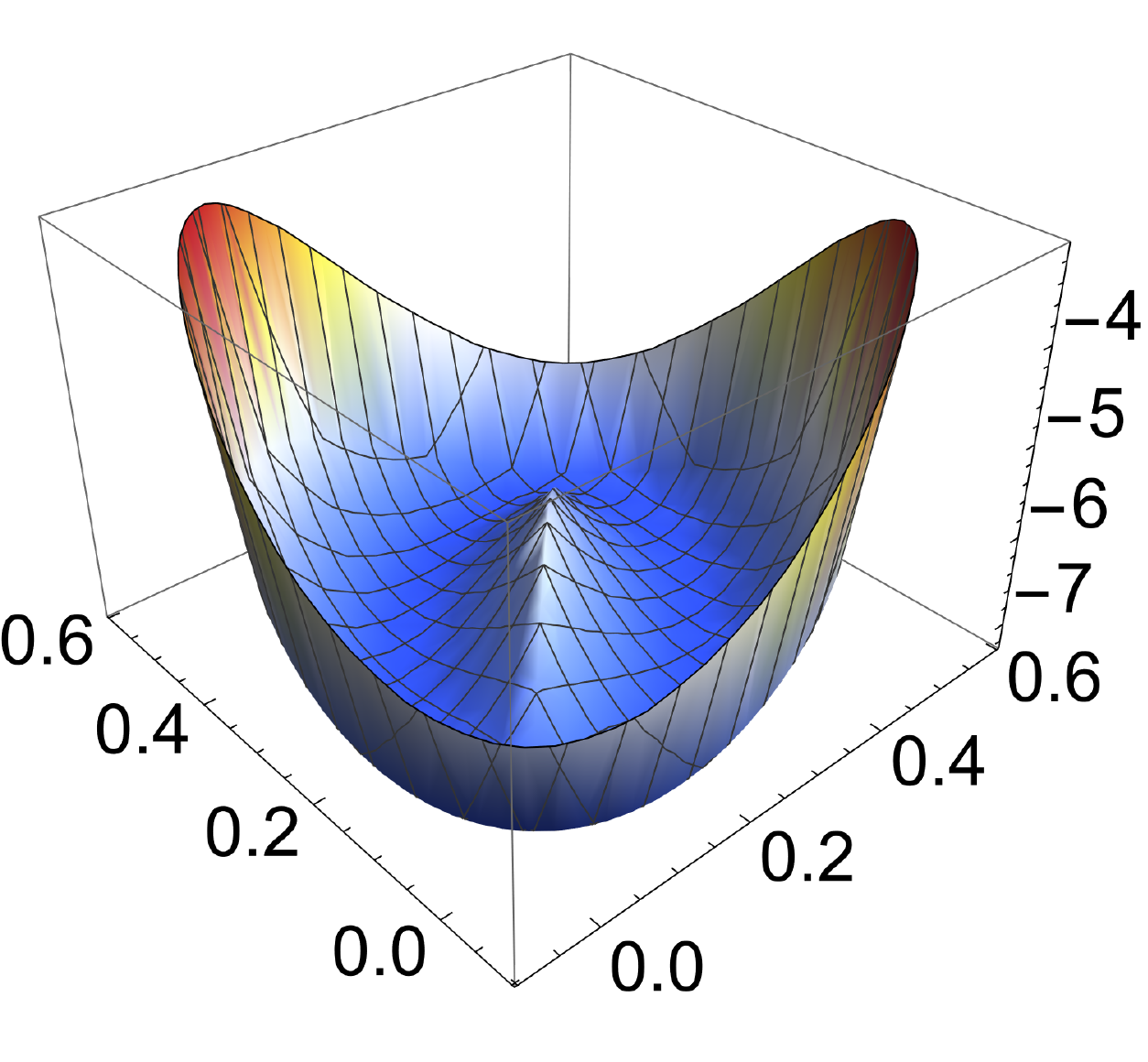}};
    \node[left=of img, node distance=0cm, anchor=center, xshift=2.78cm,yshift=-2cm,font=\color{black}] {\color{red}$\gamma_{lr}$};
\node[left=of img, node distance=0cm, anchor=center, xshift=5.5cm,yshift=-1.2cm,font=\color{black}] {\color{red}$\omega_{l,1}$};
  \node[left=of img, node distance=0cm, anchor=center, xshift=1.1cm,yshift=1.05cm,font=\color{black}] {\color{red}$\gamma_{ll}$};
  \node[left=of img, node distance=0cm, anchor=center, xshift=8.8cm,yshift=-1.2cm,font=\color{black}] {\color{red}$\omega_{l,0}$};
\node[left=of img, node distance=0cm, anchor=center, xshift=2.9cm,yshift=3cm,font=\color{black}] {\bf 1-RDMFT};
\node[left=of img, node distance=0cm, anchor=center, xshift=7.2cm,yshift=3cm,font=\color{black}] {\bf Wig\-ner quasi-DFT};
\draw[->, line width=0.4mm,red] (-0.8,2.2) -- (0.8,2.2);
\node[left=of img, node distance=0cm, anchor=center, xshift=3cm,yshift=2.2cm,font=\color{red}] {\Large $F_V[\gamma]$};
\node[left=of img, node distance=0cm, anchor=center, xshift=7.2cm,yshift=2.2cm,font=\color{red}] {\Large $\mathcal{F}_V[\omega]$};
  \node[left=of img, node distance=0cm, anchor=center, xshift=5cm,yshift=1.4cm,font=\color{black}] {$u_1=-1, u_2=-2, u_3=2$};
 \end{tikzpicture}
 \begin{tikzpicture}
 \node (img) {
  \includegraphics[scale=0.3]{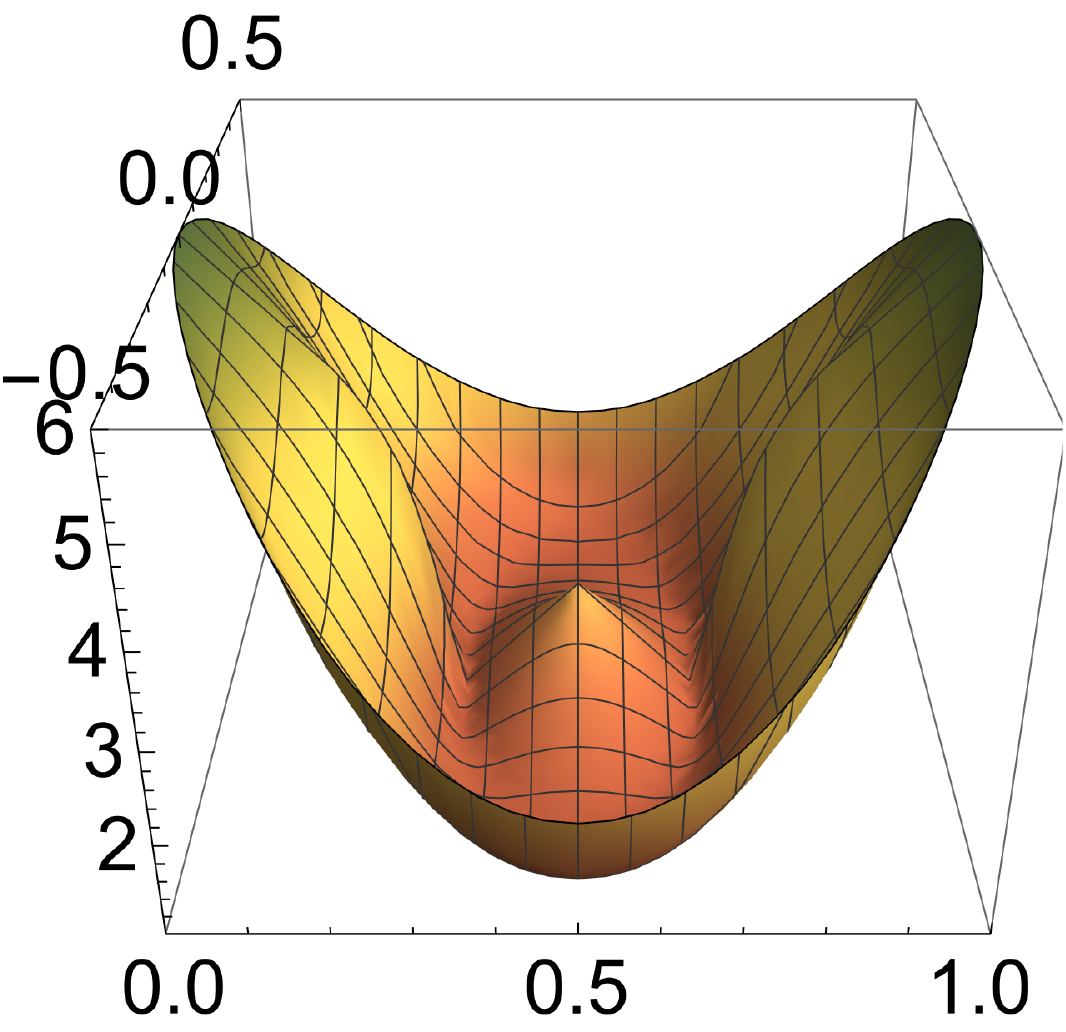}\hspace{0.7cm}
  \includegraphics[scale=0.32]{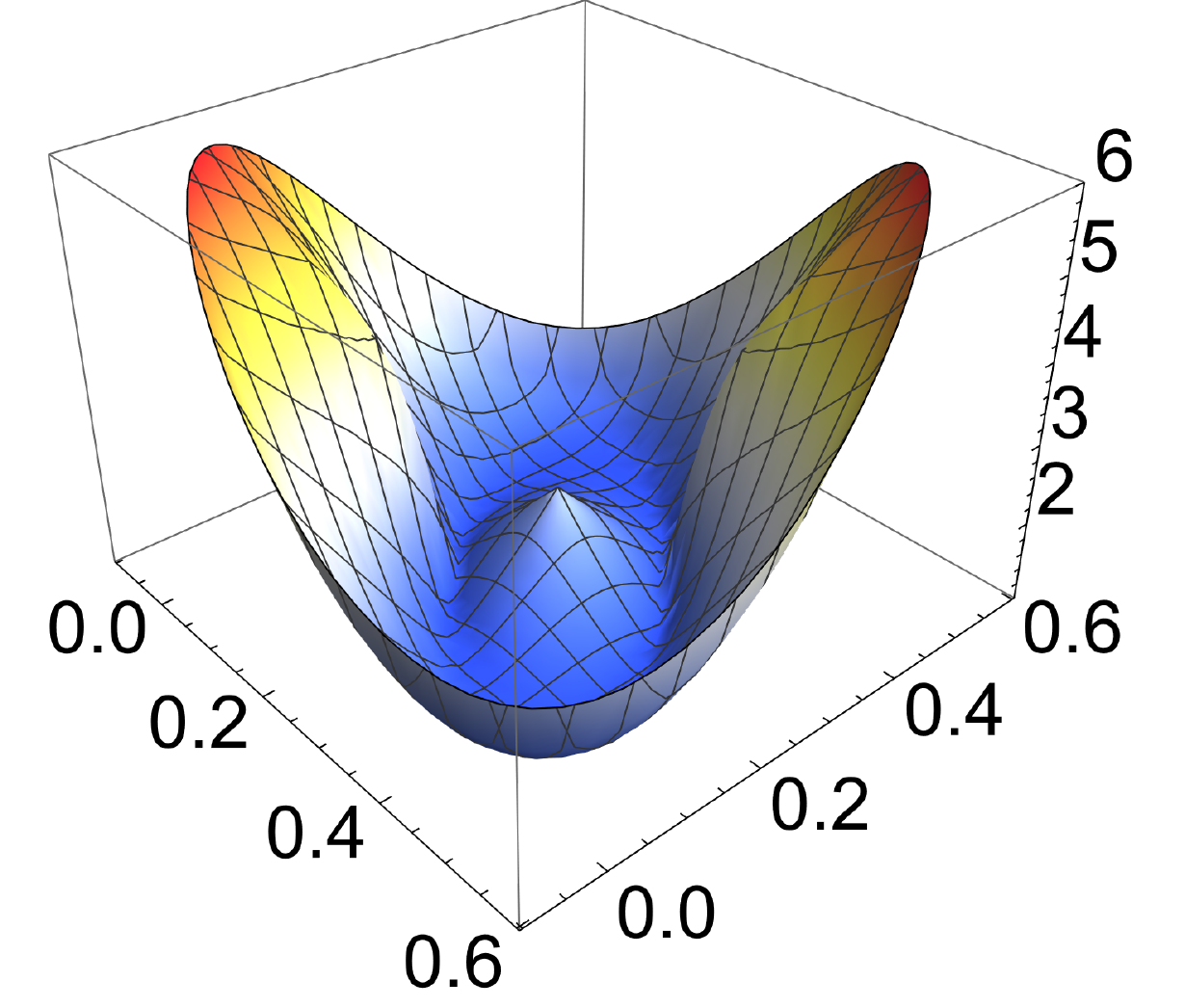}};
  \node[left=of img, node distance=0cm, anchor=center, xshift=2.85cm,yshift=-1.9cm,font=\color{black}] {\color{red} $\gamma_{ll}$};
\node[left=of img, node distance=0cm, anchor=center, xshift=5.5cm,yshift=-1.2cm,font=\color{black}] {\color{red}$\omega_{l,1}$};
  \node[left=of img, node distance=0cm, anchor=center, xshift=1.1cm,yshift=1cm,font=\color{black}] {\color{red}$\gamma_{lr}$};
  \node[left=of img, node distance=0cm, anchor=center, xshift=8.8cm,yshift=-1.2cm,font=\color{black}] {\color{red}$\omega_{l,0}$};
  \node[left=of img, node distance=0cm, anchor=center, xshift=5cm,yshift=1.4cm,font=\color{black}] {$u_1=1, u_2=0, u_3=0.5$};
 \end{tikzpicture}
\caption{Universal functionals of 1-RDMFT $F_V[\gamma]$ and Wig\-ner quasi-DFT $\mathcal{F}_V[\gamma]$ for two realizations of the generalized Bose-Hubbard dimer \eqref{hamiltletter} for three particles.}
\label{fig1}
\end{figure}
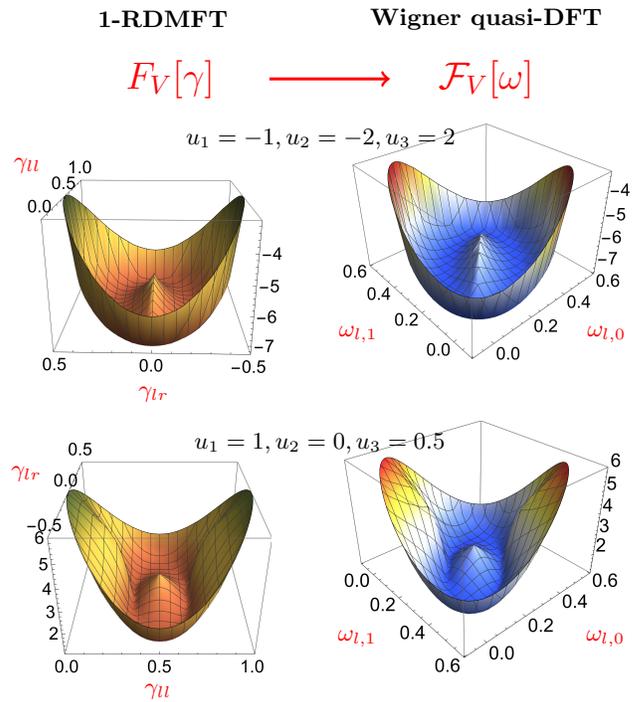

To write the corresponding (discrete) Wig\-ner transformation we follow Refs.~\cite{Feynamnn,WOOTTERS19871,PhysRevX.5.011022,doi:10.1063/1.5008653} where the Wig\-ner function is represented on a grid of twice the dimension of the underlying Hilbert space $\{j,n\}$. For the momentum basis, we choose the one in which the hopping term of the Hubbard Hamiltonian is diagonal: $\ket{n} = [\ket{l}+(-1)^n\ket{r}]/\sqrt{2}$ for $n\in\{0,1\}$. 

The 1-body Wig\-ner quasi-density can now be com\-pu\-ted: 
\begin{align}
\omega_{j,n} = \frac12 [\gamma_{jj} +(-1)^n \gamma_{lr}]\,.
\end{align}
As it should be, the marginal densities are recovered by the partial sums: $\sum_{n} \omega_{j,n} = \gamma_{jj}$ and $\sum_{j} \omega_{j,n} = \tilde{\gamma}_{nn}$, where $\tilde{\gamma}_{nn}$ is the momentum density. Since $\omega_{r,1}=\tfrac12 -\omega_{l,0}$ and $\omega_{r,0}=\tfrac12 -\omega_{l,1}$, we take $\omega_{l,0}$ and  $\omega_{l,1}$ as our two degrees of freedom. It is straightforward to check that the representability condition reads:
\begin{align}
    \left(\omega_{l,0}-\frac14\right)^2 +  \left(\omega_{l,1}-\frac14\right)^2 \leq \frac18\,.
\end{align}
Since this is the equation of a disk of radius $1/\sqrt{8}$ centered in $(\tfrac14,\tfrac14)$, one can parameterize the discrete Wig\-ner function with a radius and an angle, namely, $\omega_{l,0}(R,\phi) = \tfrac14[1 + \sqrt{2}R \cos(\phi)]$ and $\omega_{l,1}(R,\phi) = \tfrac14[1 + \sqrt{2} R \sin(\phi)]$. In Fig.~\ref{fig1} are presented two different realizations of the Ha\-mil\-to\-nian \eqref{hamiltletter} for both 1-RDMFT and quasi-DFT. It can be seen that the functional of quasi-DFT results from the respective 1-RDMFT functional after a translation, a contraction, and a rotation of 45°. This result seems to be general for lattice systems, as indicated in Appendix \ref{app1}. After applying Eq.~\eqref{eq3} (or a discrete version of \eqref{eq.crucial}) one can find $\omega_{\rm gs}$ for specific values of $t$ (the strength of the hopping term) and $v_l-v_r$ (the external potential).

\begin{figure}[!t]
\begin{tikzpicture}
 \node (img) {
 \hspace{-1.2cm}
  \includegraphics[scale=0.18]{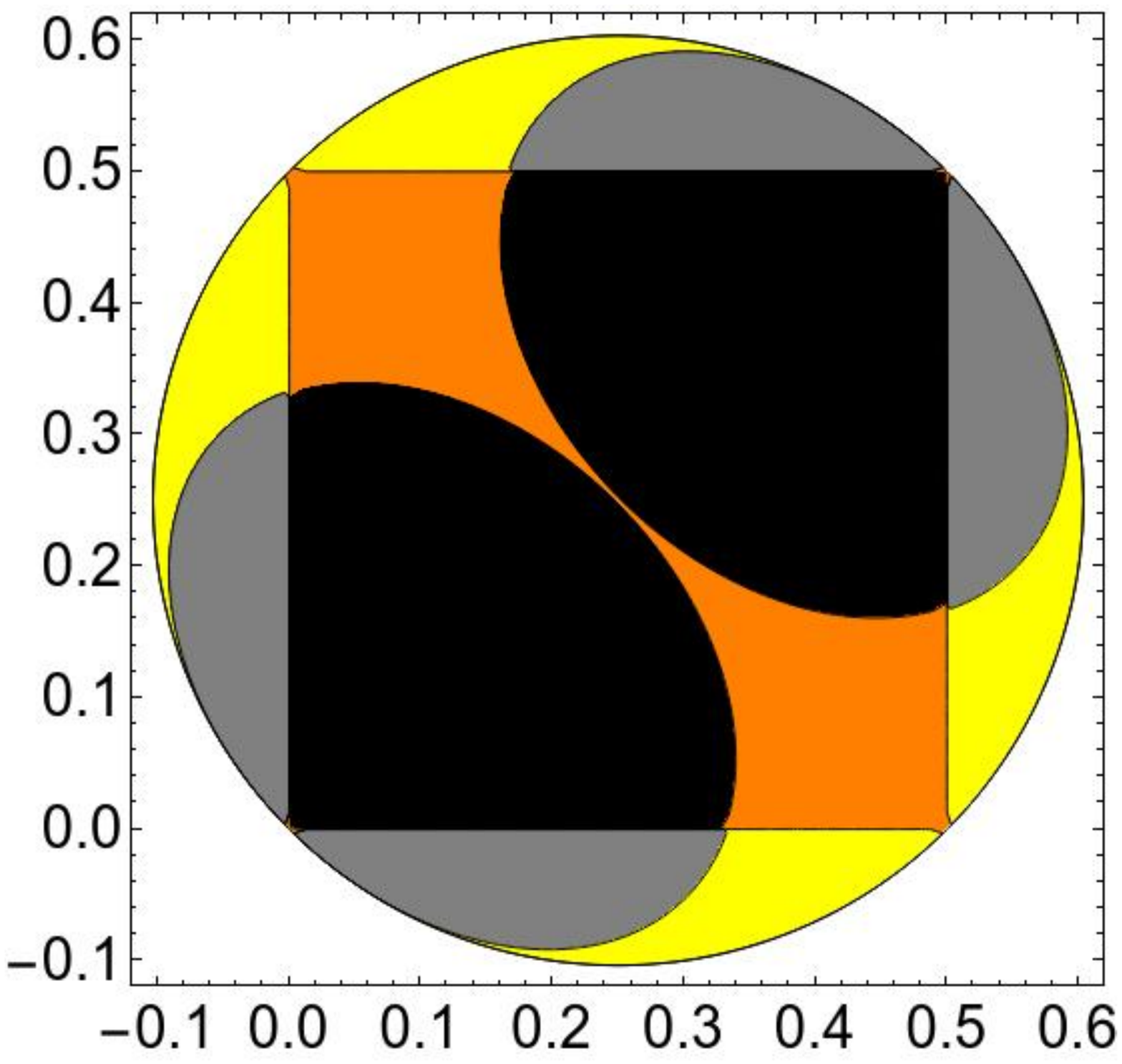}  \hspace{-1.5cm}
  \includegraphics[scale=0.18]{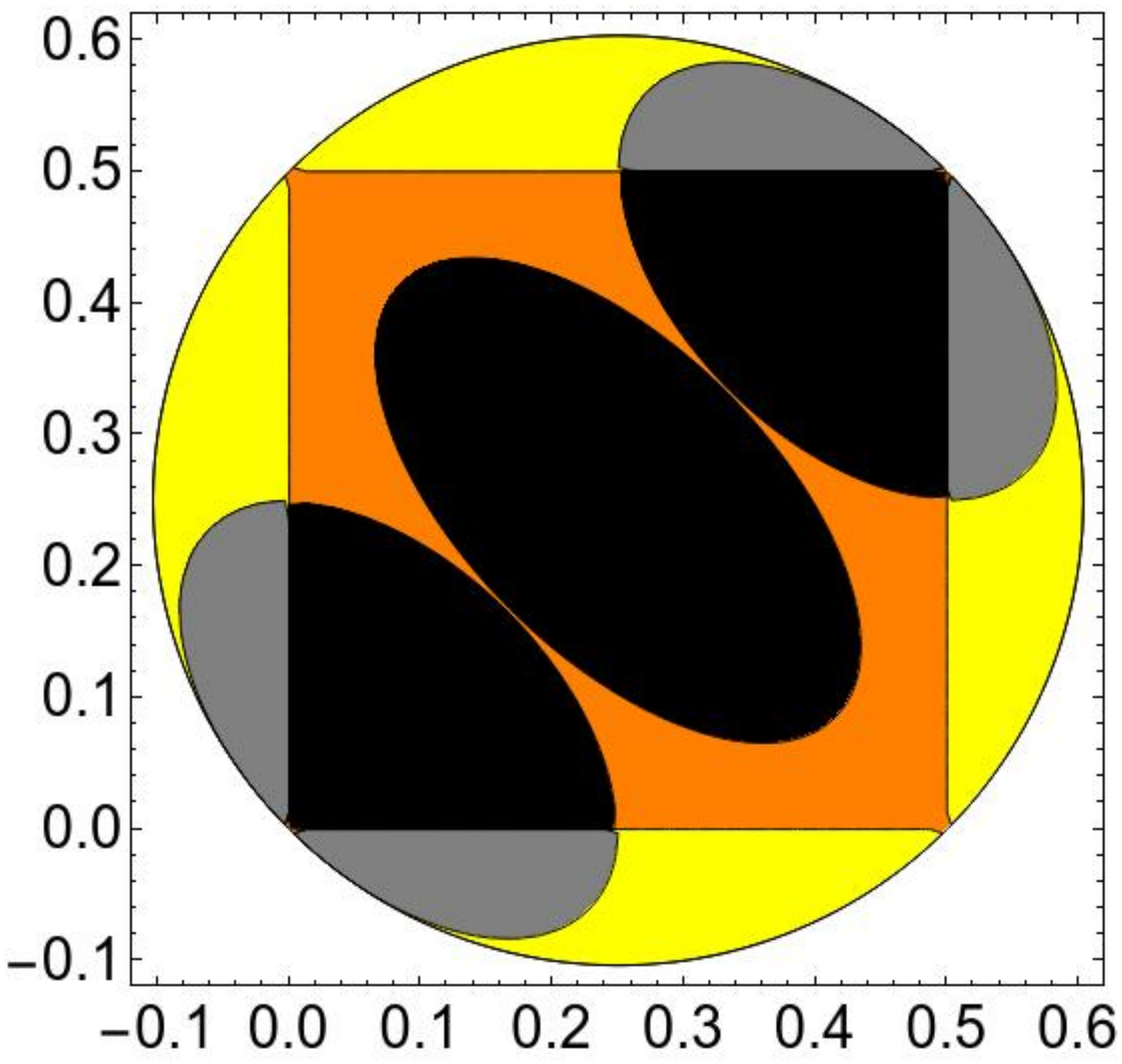}};
    \node[left=of img, node distance=0cm, anchor=center, xshift=2.9cm,yshift=-2cm,font=\color{black}] {$\omega_{l,0}$};
    \node[left=of img, node distance=0cm, anchor=center, xshift=7.1cm,yshift=-2cm,font=\color{black}] {$\omega_{l,0}$};
        \node[left=of img, node distance=0cm, anchor=center, xshift=0.7cm,yshift=-0.1cm,font=\color{black}] {$\omega_{l,1}$};
                    \node[left=of img, node distance=0cm, anchor=center, xshift=2.9cm,yshift=2cm,font=\color{black}] {$N=2$};
    \node[left=of img, node distance=0cm, anchor=center, xshift=6.9cm,yshift=2cm,font=\color{black}] {$N=3$};
    \draw[->, line width=0.2mm,red] (0.6,-0.2) -- (0,1.8);
    \draw[->, line width=0.2mm,red] (-1.1,0.3) -- (-0.2,1.8);
\node[left=of img, node distance=0cm, anchor=center, xshift=5.2cm,yshift=2.3cm,font=\color{red}] {\footnotesize{\textit{Non $v$-rep}}};
\node[left=of img, node distance=0cm, anchor=center, xshift=5.2cm,yshift=2cm,font=\color{red}] {\footnotesize{\textit{Wig\-ner neg}}};
\node[left=of img, node distance=0cm, anchor=center, xshift=4.9cm,yshift=-2.1cm,font=\color{black}] {\footnotesize{\textit{$v$-rep}}};
\node[left=of img, node distance=0cm, anchor=center, xshift=4.9cm,yshift=-2.4cm,font=\color{black}] {\footnotesize{\textit{Wig\-ner neg}}};   
 \draw[->, line width=0.2mm,black] (2,-1.2) -- (-0.15,-1.8);
 \draw[->, line width=0.2mm,black] (-2,-1.2) -- (-0.6,-1.8);
 \end{tikzpicture}
\caption{Representation of the domain of Wig\-ner 1-body quasi-densities for the Bose-Hubbard dimer \eqref{hamilt} with $u_1=1$, $u_2=u_3=0$, for 2 and 3 bo\-sons. Wig\-ner positive $\omega >0$ are represented in black (non $v$-representable) and orange ($v$-re\-pre\-sen\-ta\-ble). Wig\-ner negative $\omega$ are represented in yellow ($v$-representable) and gray (non $v$-representable).}
\label{fig2}
\end{figure}

\textit{Negativity \& $v$-representability.---} Quite remarkably, the qua\-si-DFT presented here can relate two important concepts in Wig\-ner and functional theories:  \textit{Wig\-ner negativity} and \textit{$v$-re\-pre\-sen\-tability}: Which Wig\-ner-negative 1-body quasi-densities come from ground states? We answer explicitly this question for 2 and 3 bosons for the standard Bose-Hubbard dimer in Fig.~\ref{fig2}: There are 4 disconnected ground-state regions of Wig\-ner ne\-ga\-ti\-vity! Relating these two important concepts seems to be new in the literature.

\textit{Conclusion.---} Unveiling the role of quantum effects at the classical level is a crucial problem for developing quantum technologies. The Wig\-ner quasi-probability  is usually employed as a probe of such quantumness. This letter showed that $\omega(\rr,\pp)$, the (fermionic or bosonic) 1-body Wig\-ner quasi-density, can be inserted into a func\-tio\-nal-theoretical framework in an orbital-free manner.  By providing an Euler-Lagrange equation and a Wig\-ner-Moyal eigen-equation, we showed that $\omega(\rr,\pp)$ can be computed as a stationary point without referring to orbital equations, circumventing some known problems of orbital-free DFT (e.g., approximating the kinetic energy functional or finding the Pauli potential). In this framework, the concepts of Wigner negativity and $v$-representability can be related. We would like to emphasize that one of the most important aspects of our results is that the $\star$-product gives a very rich structure for extracting the corresponding 1-particle Wigner function. In that sense, quasi-DFT is a functional theory that can connect directly with DFT and with semiclassical expansions of the many-body problem. There are several potential research directions from our results: First, one could develop machine learning quasi-DFT functionals which is now current practice for standard DFT \cite{Grisafi2022,sciencepaper,doi:10.1126/sciadv.abq0279,Gedeon_2022}. Second, since Wig\-ner negativities carry important quantum information it will be interesting to see what information they can unveil for electronic correlations \cite{Izsak2023,Zhang2021} or fermionic entanglement  \cite{PhysRevLett.130.150201}. Finally, it could be quite promising to tackle ---within this orbital-free framework--- quantum excitations in the same spirit of state-average calculations \cite{PhysRevLett.129.066401} or the recently formulated $\mathbf{w}$-1-RDMFT \cite{PhysRevLett.127.023001}.

\vspace{2cm}
\begin{acknowledgments}
\vspace{-0.2cm}
I gratefully thank Luis Colmenarez, Julia Liebert, Eli Kraisler, and Jeff Maki for insightful discussions and for providing important comments on the manuscript, and acknowledge the European Union’s Horizon Europe Re\-search and Innovation program  un\-der the Marie Skło\-dowska-Curie grant agreement n°101065295. I also thank Ana Maria Rey and the warm atmosphere of her group at JILA where this paper took its final shape.   
\end{acknowledgments}

\onecolumngrid

\appendix

\section{Discrete Wigner formalism for the Hubbard model}
\label{app1}

Here we apply the discrete Wigner formalism to the Hubbard model of $L$ sites. This is defined in the $L$-dimensional Hilbert space $\mathcal{H}^L$ whose position basis is $\mathcal{S}=\{\ket{1},...,\ket{L} \}$. Another orthonormal basis for the same Hilbert space is $\{\ket{\phi_0},...,\ket{\phi_{L-1}} \}$, defined by the Fourier transform: 
\begin{align}
    \ket{\phi_m} = \frac{1}{\sqrt{L}}\sum^L_{n=1} e^{in\phi_m} \ket{n}\,,
\end{align}
with $\phi_m = \frac{2\pi}{L} m$. The set of pairs $\{n,\phi_m\}_{n,m}$ constitutes a $L\times L$ grid. This is the phase space $\Gamma^L$ associated with the Hilbert space $\mathcal{H}^L$ \cite{WOOTTERS19871}.

The operators $\hat n = \sum_n n \ket{n}\bra{n}$ and $\hat\phi = \sum_m \phi_m \ket{\phi_m}\bra{\phi_m}$ can be used to construct the following unitary operators:
\begin{align}
    \hat V = \exp\left(i\frac{2\pi}{L}\hat n\right) \qquad {\rm and} \qquad   \hat U = \exp(i\hat \phi)\,,
\end{align}
which satisfy the Weyl relation \cite{doi:10.1063/1.5008653}:
\begin{align}
\hat D(k,l) \equiv \exp\left(-i\frac{\pi kl}{L}\right)\hat U^k\hat V^l =
\exp\left(i\frac{\pi kl}{L}\right)\hat U^l\hat V^k \,, \qquad k,l\in   \mathbb{Z}\,.  
\end{align}
With this operator, the authors of Ref.~\cite{doi:10.1063/1.5008653} define the phase-space point operator:
\begin{align}
    \hat \Omega_\kappa(n,\phi_m) = \frac{1}{L} \sum_{k,l} \kappa(k,l) \hat{D}(k,l) \exp\left[-i\left(k\phi_m+\frac{2\pi}{L}ln\right)\right]\,,
\end{align}
with a kernel $\kappa(k,l)$, whose properties are determined by the properties of $\hat \Omega_\kappa$. In particular, the operator's hermiticity condition: $\hat \Omega_\kappa(n,\phi_m)  = \hat \Omega^\dagger_\kappa(n,\phi_m)$, $\forall (n,\phi_m) \in \Gamma^L$, which is needed to map phase-space functions to hermitian operators, results in the condition $\kappa^*(k,l) = (-1)^{L+k+l} \kappa(L-k,L-l)$. For odd $L = 2N +1$, for instance, a kernel can be chosen to be \cite{doi:10.1063/1.5008653}: $\kappa(k,l) = \cos(\pi kl/L)$. 

The map between $f(n,\phi_m)$, a real function in $\Gamma^L$, and $\hat f$, an operator in $\mathcal{H}^L$, is realized by means of the following relation:
\begin{align}
\hat f = \frac{1}{L} \sum_{m,n} f(n,\phi_m)  \hat \Omega_\kappa (n,\phi_m)\,.
\label{eqa1}
\end{align}
Eq.~\eqref{eqa1} can now be used to find the Wigner quasi-distribution. Since the average value of the
observable represented by the operator $\hat f$ in a state defined by the density operator $\hat \gamma$ reads
\begin{align}
    \Tr[\hat \gamma \hat f ] = \frac{1}{L} \sum_{m,n} f(n,\phi_m)  \Tr \left[\hat \gamma \hat \Omega_\kappa (n,\phi_m)\right]\,.
\end{align}
a natural definition for the Wigner quasi-probability (for the kernel $\kappa$) arises:
$\omega(n,\phi_m) = \Tr \left[\hat \gamma \hat \Omega_\kappa (n,\phi_m)\right]$. From this definition, one can write:
\begin{align}
\omega(n,\phi_m)  = \sum_{m',n'}\mathcal{D}(n,\phi_m;n',m')
\bra{n'} \hat \gamma\ket{m'} \,,
\label{eqA1}
\end{align}
where
$\mathcal{D}(n,\phi_m;n',m')  = 
 \frac{1}{L^2} \sum_{k,l,s}
\kappa(k,l) \exp\left[i \left(\frac{\pi kl}{L} + k\phi_s+n'\phi_{s+l}-m'\phi_s-k\phi_m-\frac{2\pi}{L}ln\right)\right]$.
If one vectorizes both $\omega$ and $\gamma$, to wit, 
\begin{align}
    \kett{\omega} = \begin{pmatrix}
        \omega(1,\phi_0) \\
        \omega(1,\phi_1)\\ 
        \omega(1,\phi_2) \\
        \vdots
    \end{pmatrix} \qquad {\rm and} \qquad     \kett{\gamma} = \begin{pmatrix}
        \bra{1}\gamma\ket{1} \\
        \bra{1}\gamma\ket{2} \\ 
        \bra{1}\gamma\ket{3} \\
        \vdots
    \end{pmatrix}\,,
\end{align}
one can formally write \eqref{eqA1} as $\kett{\omega} = \hat{\mathcal{D}}  \kett{\gamma}$.

\section{The Generalized Bose-Hubbard dimer}

In this section, we focus on the generalized Bose-Hubbard dimer, whose Hamiltonian reads
\begin{equation}
H = -t (b_l^\dagger b_r+b_r^\dagger b_l) + \!\sum_{j=l/r}\! v_j \hat{n}_j + V\,, 
\label{hamilt}
\end{equation}
with 
\begin{equation}
V = u_1 \left[\hat{n}_l(\hat{n}_l-1) + \hat{n}_r(\hat{n}_r-1) \right]+ u_2 \hat{n}_l \hat{n}_r + u_3 \left[(b^\dagger_{l})^2 (b_{r})^2+(b^\dagger_{r})^2 (b_{l})^2\right]\,.
\label{hamilt2}
\end{equation}
The operators $b^\dagger_{j}$ and $b_{j}$ create and annihilate a particle on the sites $j=l/r$, and $\hat n_{j}$ is the corresponding particle-number ope\-ra\-tor. Any $N$-body ground state of the Hamiltonian \eqref{hamilt} can be expressed as a linear combination of the configuration states $\ket{n,N-n}$. Assuming real wave functions, we represent the 1-RDM $\gamma \equiv \mbox{Tr}_{N-1}[\Gamma]$ of any pure or ensemble state with respect to the lattice site states $\ket{l},\ket{r}$,
\begin{align}
\gamma_{ij} \equiv \frac1{N}\bra{\Psi}  b_i^\dagger  b_j \ket{\Psi}\,,\qquad i,j=l,r\,.
\end{align}
Since $\gamma_{ll} + \gamma_{rr} = 1$ (by normalization) and
$\gamma_{lr} = \gamma_{rl}$ the 1RDM is fully determined by two free parameters. We represent the 1-RDM as: 
\begin{align}
    \gamma = \begin{pmatrix}
\gamma_{ll} & \gamma_{lr}\\
\gamma_{lr} & 1-\gamma_{ll}
\end{pmatrix} \,.
\end{align}
The only two degrees of freedom of this matrix can be represented in a vector form:
$\kett{\gamma} = \begin{pmatrix}
        \gamma_{ll} \\
        \gamma_{lr} 
    \end{pmatrix}$.

\subsection{1-body Wigner function}
Our goal is to find the Wigner function associated with this matrix in the grid $\{(l,0),(l,1),(r,0),(r,1)\}$, with the momentum basis: $\ket{0} \equiv \frac1{\sqrt{2}}(\ket{L}+\ket{R})$ and $\ket{1} \equiv \frac1{\sqrt{2}}(\ket{L}-\ket{R})$. This grid can be seen as a two-dimensional vector space over a finite field, in which the Wigner function is defined:
\[ 
\begin{array}{c@{}c@{}c}
    \begin{matrix}
       &  &   | & & \\
       & \omega(l,1) & |   & \omega(r,1) & \\
       &  &   | & &  
   \end{matrix} \\
   \hline
    \begin{matrix}
       &  &   | & & \\
       & \omega(l,0) & |   & \omega(r,0) & \\
       &  &   | & &  
   \end{matrix}
\end{array} 
\]

Notice that $\gamma$ can be written as
\begin{align}
    \gamma = \left(\frac12+\vec{\gamma}\cdot \vec{\sigma}\right)\,,
\end{align}
where $\vec{\gamma} = (\gamma_{lr},0,\gamma_{ll}-\tfrac12)$
and $\vec{\sigma}= (\sigma_{x},\sigma_y,\sigma_{z})$ are the Pauli matrices. For a qubit a phase-space point operator is known to be \cite{doi:10.1063/1.5008653}:
\begin{align}
\Omega(n,\phi_m) = \frac12 \left[1 + (-1)^m (\ket{0}\bra{0})-\ket{1}\bra{1})+
(-1)^n (\ket{0}\bra{1})+\ket{1}\bra{0})+ i (-1)^{n+m} (\ket{0}\bra{1})-\ket{1}\bra{0})\right]\,.
\end{align}
Therefore, by computing $\omega(n,\phi_m) = \frac14 \Tr[\gamma  \Omega(n,\phi_m)]$ one finds in vectorized form the following equation:
\begin{align}
    \kett{\omega(\gamma)} = \tfrac12 \mathcal{D} \kett{\gamma}\,,
    \label{eqw}
\end{align}
where 
\begin{align}
   \kett{\omega} =
  \begin{pmatrix}
        \omega(l,0) \\
        \omega(l,1) 
    \end{pmatrix}  \qquad {\rm and } \qquad \mathcal{D} = 
        \begin{pmatrix}
        1 & 1  \\
        1 & -1 
    \end{pmatrix}
\end{align}
with $\omega(r,1) = \tfrac12 -\omega(l,0)$ and $\omega(r,0) = \tfrac12 -\omega(l,1)$. Notice that $\mathcal{D}$ is an orthogonal  matrix. Inverting \eqref{eqw} one gets $ \kett{\gamma(\omega)} = \mathcal{D}  \kett{\omega}$.

\subsection{Representability}

Since $ \gamma^2 \leq  \gamma$, the domain  of pure/ensemble $N$-representable 1RDMs takes the form of a disc of radius $\tfrac12$ given by
\begin{equation}
\left(\gamma_{ll} - \frac{1}2 \right)^2 + \gamma_{lr}^2 \leq \frac{1}4.
\label{eqrep}
\end{equation}
The area of this disc is $A_N = \pi /4$ and its boundary $\partial\mathcal{P}_p$ (i.e.,
$\gamma_{LR}^2 + (\gamma_{LL}-\tfrac12)^2 = \tfrac14$) corresponds to complete Bose-Einstein condensation (BEC) \cite{PhysRevLett.124.180603}. Plugging \eqref{eqw} in \eqref{eqrep} one gets:
\begin{equation}
\left(\omega_{l,0} - \frac{1}4 \right)^2 +\left(\omega_{l,1} - \frac{1}4 \right)^2 \leq \frac{1}8.
\label{eqrep2}
\end{equation}

\section{Proof of Eq.~(7)}
\label{app3}

We take the the functional derivative of $\mathcal{E}[\omega] - \mu N[\omega]$ at the point $\omega$ in the direction of $\omega$: 
\begin{align}
   0 = \delta \big(\mathcal{E}[\omega] -\mu N [\omega]\big) &= \int \frac{\delta \big(\mathcal{E}[\omega] -\mu N [\omega]\big)}{\delta \omega(\rr,\pp)} \omega(\rr,\pp) d\Omega \nonumber \\
    & =\int \left[h(\rr,\pp)+ \frac{\delta \mathcal{F}[\omega]}{\delta \omega(\rr,\pp)} -\mu \right]\omega(\rr,\pp)d\Omega \nonumber \\
    & =\int \left[h(\rr,\pp) \omega(\rr,\pp)+ \frac{\delta \mathcal{F}[\omega]}{\delta \omega(\rr,\pp)} \omega(\rr,\pp) -\mu  \,\omega(\rr,\pp) \right]d\Omega  \nonumber \\
     & =\int \left[h(\rr,\pp) \star \omega(\rr,\pp)+ \frac{\delta \mathcal{F}[\omega]}{\delta \omega(\rr,\pp)} \star \omega(\rr,\pp) -\mu \, \omega(\rr,\pp) \right]d\Omega \,,
\end{align}
where $d\Omega = d^3\rr d^3\pp$. In the last line we have used the fact that $\int f g d\Omega = \int f \star g d\Omega$. Therefore, we conclude that at each phase-space point:
\begin{align}
    h(\rr,\pp) \star \omega(\rr,\pp)+ \frac{\delta \mathcal{F}[\omega]}{\delta \omega(\rr,\pp)} \star \omega(\rr,\pp) -\mu \, \omega(\rr,\pp) = 0\,.
\end{align}
Since  $\int f g d\Omega = \int g \star f d\Omega$ also holds, we also conclude that:
\begin{align}
    \omega(\rr,\pp) \star h(\rr,\pp) + \omega(\rr,\pp) \star  \frac{\delta \mathcal{F}[\omega]}{\delta \omega(\rr,\pp)}  -\mu \, \omega(\rr,\pp) = 0\,.
\end{align}

\section{Hartree-Fock in phase space}

In this last section, we investigate the form of the orbital-free Hartree-Fock equations in phase space for a system of $N$ electrons.  As the respective wave function is a single Slater determinant, the 1-body reduced-density matrix is a projector: 
$$
\gamma(\rr,\rr') = \sum^N_{n=1} \varphi_n(\rr)\varphi^*_n(\rr')\,,
$$
with $\int \varphi_n(\rr)\varphi_m^*(\rr)d^3\rr = \delta_{nm}$. The first result we will prove is that the corresponding Wigner function satisfies $\omega \star \omega = \omega$.
\begin{proof}
Let us first define the Wigner phase-space orbitals $\chi_n(\rr,\pp) = \int \varphi_n(\rr-\zz)\varphi_n^*(\rr+\zz) e^{2i \pp\cdot \zz}d^3\zz$. They satisfy the following equation:
\begin{align*}
& \chi_n(\rr,\pp) \star \chi_m(\rr,\pp) \\ &=
\int \, \chi_n(\rr',\pp') \,\chi_m(\rr'',\pp'')
e^{2i(\rr\cdot \pp'-\rr'\cdot \pp + \rr'\cdot \pp'' -\rr''\cdot \pp'+ \rr''\cdot \pp-\rr\cdot \pp'')}\,
d\Omega' \, d\Omega'' \\&= \int \varphi_n(\rr'-\zz')\, \varphi^*_n(\rr'+\zz') \,
\varphi_m(\rr''-\zz'')\, \varphi^*_m(\rr''+\zz'') \, e^{2i (\rr'' - \rr') \cdot \pp} 
e^{2i (\zz' + \rr - \rr'') \cdot \pp'} e^{2i (\zz'' + \rr' - \rr) \cdot \pp''}
d\Omega' \, d\Omega'' \, d^3\zz'\, d^3\zz '' \\
&= \int \varphi_n(\rr'-\zz')\, \varphi^*_n(\rr'+\zz') \,
\varphi_m(\rr''-\zz'')\, \varphi^*_m(\rr''+\zz'')  e^{2i (\rr'' - \rr') \cdot \pp}\,
\delta(\zz' + \rr - \rr'') \, \delta (\zz'' + \rr' - \rr) 
\, d^3\rr' \, d^3\rr'' \, d^3\zz' \, d^3\zz ''  \\
&=  \int \varphi_n(\rr'+\rr-\rr'')\, \varphi^*_n(\rr'-\rr+\rr'') \,
\varphi_m(\rr''+\rr'-\rr)\, \varphi^*_m(\rr''-\rr'+\rr)  e^{2i (\rr'' - \rr') \cdot \pp} \,d^3\rr' d^3\rr'' .
\end{align*}
Letting $\mathbf{u} = \rr''-\rr'$ and $\mathbf{v} = \rr'+\rr''-\rr$, we have:
\begin{align*}
\chi_{n}(\rr,\pp) \star \chi_{m}(\rr,\pp) &=
\int \varphi_n(\rr - \mathbf{u})\, \varphi^*_m(\rr+\mathbf{u}) \,
e^{2i \mathbf{u} \cdot\pp} d^3\mathbf{u} \int
\varphi_m(\mathbf{v})\, \varphi^*_n(\mathbf{v}) \, d^3\mathbf{v} = \delta_{nm}\chi_n(\rr,\pp)\,.
\end{align*}
As a consequence, 
\begin{align}
    \omega(\rr,\pp) \star \omega(\rr,\pp) = \sum_{nm} \chi_n(\rr,\pp) \star \chi_m(\rr,\pp) = \sum_{nm} \chi_n(\rr,\pp) \delta_{nm} = \omega(\rr,\pp)\,.
\end{align}
\end{proof}
This result indicates that we have to solve the Hartree-Fock functional subject to the condition $\omega\star\omega = \omega$ and the normalization $\int \omega(\rr, \pp) d\Omega =N$. Using the Lagrange multipliers $\omega(\rr,\pp)$ and $\beta$, the variational problem reads:
\begin{align}
    \delta\left\{\mathcal{E}_{\rm HF}[\omega] - \int \alpha(\rr,\pp)\left[\omega (\rr,\pp)\star \omega(\rr,\pp)-\omega(\rr,\pp)\right]\,d\Omega - \beta \left[\int \omega(\rr,\pp) \,d\Omega -N \right]\right\} = 0\,.
    \label{eqHF}
\end{align}
Before performing the variation note that
\begin{align}
& \frac{\delta}{\delta \omega(\rr,\pp)} 
\int \alpha(\rr,\pp) \,\omega (\rr,\pp)\star \omega(\rr,\pp)\,d\Omega \nonumber  \\ &\qquad = \frac{\delta}{\delta \omega(\rr,\pp)}  \int \alpha(\rr,\pp) \, \omega(\rr',\pp') \,\omega(\rr'',\pp'') e^{2i(\rr\cdot \pp'-\rr'\cdot \pp + \rr'\cdot \pp'' -\rr''\cdot \pp'+ \rr''\cdot \pp-\rr\cdot \pp'')}\, d\Omega\, d\Omega' \, d\Omega'' \nonumber  \\ &\qquad =  \int \alpha(\rr',\pp') \,\omega(\rr'',\pp'') e^{2i(\rr'\cdot \pp-\rr\cdot \pp' + \rr\cdot \pp'' -\rr''\cdot \pp+ \rr''\cdot \pp'-\rr'\cdot \pp'')}\,  d\Omega' \, d\Omega'' \nonumber \\
&\qquad +  \int \alpha(\rr'',\pp'') \,\omega(\rr',\pp') e^{2i(\rr''\cdot \pp'-\rr'\cdot \pp'' + \rr'\cdot \pp -\rr\cdot \pp'+ \rr\cdot \pp''-\rr''\cdot \pp)}\,  d\Omega' \, d\Omega''  \nonumber \\
&\qquad = \omega(\rr,\pp)\star \alpha(\rr,\pp) + \alpha(\rr,\pp)\star \omega(\rr,\pp)\,.
\end{align}
Using this result in Eq.~\eqref{eqHF} we obtain
\begin{align}
  L(\rr,\pp) \equiv   f_{\rm HF}(\rr,\pp) - \omega(\rr,\pp)\star \alpha(\rr,\pp) - \alpha(\rr,\pp)\star \omega(\rr,\pp) + \alpha(\rr,\pp) - \beta = 0 \,,
\end{align}
where $f_{\rm HF}(\rr,\pp) = \delta \mathcal{E}_{\rm HF}[\omega] /\delta \omega(\rr,\pp)$. Multiplying (with the  $\star$-product) this equation on the left  by $\omega(\rr,\pp)$ (i.e, $\omega(\rr,\pp)\star L(\rr,\pp)$) and on the right (i.e, $L(\rr,\pp) \star \omega(\rr,\pp)$), and then subtracting both equations we obtain that $\omega$ $\star$-anticommutes with $f_{\rm HF}(\rr,\pp)$:
\begin{align}
    [f_{\rm HF}(\rr,\pp) , \omega(\rr,\pp)]_\star \equiv
    f_{\rm HF}(\rr,\pp) \star \omega(\rr,\pp) - \omega(\rr,\pp) \star f_{\rm HF}(\rr,\pp) = 0 \,.
\end{align}
This is the equation of $\omega(\rr,\pp)$ within Hartree-Fock theory. Recall that it admits an expansion in $\hbar$. For this reason, this equation allows a semiclassical expansion that does not exist in the double-coordinate representation. 

To finish the calculation we give now the explicit form of $f_{\rm HF}(\rr,\pp)$. Let us define the 1-particle Hamiltonian $\pp^2/2m + v(\rr)$, with $v(\rr)$ being the external potential. Using the inverse of the Wigner transformation, the Hartree-Fock energy reads:
\begin{align*}
     \mathcal{E}_{\rm HF}[\omega]  = \int h(\rr,\pp) \, \omega (\rr,\pp) \, d\Omega + \frac12\int \frac{\omega(\rr,\pp)\,\omega(\rr',\pp')}{|\rr - \rr'|}d\Omega \,d\Omega' -\frac12\int e^{i(\pp-\pp')\cdot(\rr-\rr')} \frac{\omega((\rr+\rr')/2,\pp)\,\omega((\rr+\rr')/2,\pp')}{|\rr - \rr'|}d\Omega \,d\Omega'  \,.
\end{align*}
A straightforward calculation finally gives:
\begin{align*}
     f_{\rm HF}(\rr,\pp)  =  h(\rr,\pp)  + \int \frac{\omega(\rr',\pp')}{|\rr - \rr'|}\,d\Omega' - \int \frac{e^{i(\pp-\pp')\cdot\mathbf{\rr'}} }{|\mathbf{\rr'}|}\omega(\rr,\pp')\,d\Omega'  \,.
\end{align*}

\twocolumngrid
\bibliography{Refs2}

\end{document}